\documentclass{aa}  
\usepackage{graphicx}
\usepackage{txfonts}
\usepackage{color}
\usepackage{natbib}
\bibpunct[ ]{(}{)}{;}{a}{}{}
\begin{document}
   \title{MOST\thanks{Based on data from the MOST satellite, a Canadian Space Agency mission,
jointly operated by Dynacon Inc., the University of Toronto Institute for Aerospace Studies and the 
University of British Columbia, with the assistance of the University of Vienna.} photometry of 
the roAp star 10\,Aql}

	\titlerunning{MOST photometry of 10\,Aql}
	\authorrunning{D. Huber et al.}

   \author{D. Huber\inst{1}
		\and H. Saio\inst{2}
		\and M. Gruberbauer\inst{1}
		\and W.W. Weiss\inst{1}
		\and J.F. Rowe\inst{3} 
		\and M. Hareter\inst{1}
		\and T. Kallinger\inst{1} 
		\and P. Reegen\inst{1}  
		\and J.M. Matthews\inst{3}
		\and R. Kuschnig\inst{3}
		\and D.B. Guenther\inst{4}
		\and A.F.J. Moffat\inst{5}
		\and S. Rucinski\inst{6}
		\and D. Sasselov\inst{7} 
		\and G.A.H. Walker\inst{3}}
		
	\offprints{Daniel Huber, huber@astro.univie.ac.at}

   \institute{Institute for Astronomy (IfA), University of Vienna,
              T\"urkenschanzstrasse 17, 1180 Vienna, Austria
        \and Astronomical Institute, Graduate School of Science, Tohoku University, Sendai, 980-8578, Japan,
		\and Dept. of Physics and Astronomy, University of British Columbia, 6224 Agricultural Road, Vancouver, BC V6T 1Z1, Canada 
		\and Dept. of Astronomy and Physics, St. Mary's University, Halifax, NS B3H 3C3, Canada 
		\and D\'ept. de physique, Univ. de Montr\'eal, C.P. 6128, Succ. Centre-Ville, Montr\'eal, QC H3C 3J7, Canada 
		\and Dept. of Astronomy \& Astrophysics, David Dunlop Obs., Univ. Toronto, P.O. Box 360, Richmond Hill, ON L4C 4Y6, Canada 
		\and Harvard-Smithsonian Center for Astrophysics, 60 Garden Street, Cambridge, MA 02138, USA 
			}
             
   \date{Received; accepted}

  \abstract
   {We present 31.2 days of nearly continuous MOST photometry of the roAp star 10\,Aql.}
   {The goal was to provide an unambiguous frequency identification for this little 
   studied star, as well as to discuss the detected frequencies in the context of magnetic models and 
   analyze the influence of the magnetic field on the pulsation.}
   {Using traditional Fourier analysis techniques on three independent data reductions, intrinsic frequencies for the star 
   are identified. Theoretical non-adiabatic axisymmetric modes influenced by a magnetic field having polar field strengths 
   $B_{\mathrm{P}}$ = 0--5\,kG were computed to compare the observations to 
   theory.}
   {The high--precision data allow us to identify three definite intrinsic 
   pulsation frequencies and two other candidate frequencies with 
   low S/N. Considering the observed spacings, only one (50.95\,$\mu$Hz) is consistent 
   with the main sequence nature of roAp stars. The comparison with theoretical models yields a best fit 
   for a 1.95\,$M_{\sun}$ model having solar metallicity, suppressed envelope convection, and homogenous 
   helium abundance. Furthermore, our analysis confirms the suspected slow rotation of the star 
   and sets new lower limits to the rotation period ($P_{\rm rot}\geq 1$ month) and inclination ($i>30\pm10^{\circ}$).}
   {The observed frequency spectrum is not rich enough to unambiguously identify a model. On the other 
   hand, the models hardly represent roAp stars in detail due to the approximations needed to 
   describe the interactions of the magnetic field with stellar structure and pulsation. Consequently, 
   errors in the model frequencies needed for the fitting procedure can only be estimated. 
   Nevertheless, it is encouraging that models which suppress convection and include solar metallicity, 
   in agreement with current concepts of roAp stars, fit the observations best.}
   

   \maketitle

\section{Introduction}

	Rapidly oscillating Ap (roAp) stars are classified as chemically peculiar main sequence stars pulsating 
	with periods ranging from 6--21 mins. Since the first discovery by \citet{kurtz78}, more than 30 
	members have been identified and their pulsations are generally believed to be high-order 
	low-degree magneto-acoustic modes. The roAp stars are characterized by strong global magnetic fields with 
	polar strengths up to several kG. 
	
	The discovery of a synchronous modulation of the pulsation amplitude 
	with the magnetic field strength led to the oblique pulsator model, which 
	assumes the pulsation axis is aligned with the magnetic axis which is itself oblique to 
	the axis of rotation. This concept  naturally explains the observed frequency triplets as 
	axisymmetric dipolar modes modulated by the rotation period \citep{kurtz82}. This model was 
	challenged by \citet{bigotdziem}, who tookg into account rotational effects such as the
	centrifugal force and argued in favor of non-axisymmetric modes. Assuming an axisymmetric mode 
	aligned with the magnetic axis, \citet{saio04} found that the amplitude of the pulsation 
	is confined to the magnetic axis, which was confirmed observationally by \citet{kochukhov06a} 
	analyzing spectroscopic line profile variations of the rapidly rotating roAp star HR\,3831.
	
	The mechanism driving the pulsation in roAp stars is widely believed to be the $\kappa$-mechanism acting in the H ionization zone. Attempts have been made to explain the lack of low-order
	pulsation modes despite the vicinity to classical $\delta$\,Scuti pulsators in the Hertzsprung-Russell 
	diagram (HRD) by both indirect and direct effects of the magnetic field. With the inclusion 
	of gravitational 
	settling of helium in the outer layers of the star due to inhibited meridional circulation as well as suppression 
	of envelope convection at the poles due to the magnetic field, it became possible to drive the high-order pulsations as 
	observed \citep{balmforth}; however, some lower order modes were also excited in these models.
	
	\citet{saio05} concentrated on the direct effect of the magnetic field, showing that the decoupling of the 
 	pulsation into an acoustic wave and a magnetic slow wave (which dissipates in the interior) 
 	is sufficient to damp out all the low-order modes, while high order modes remain unstable. Beside the 
 	excitation analysis, the direct effect on the observed frequencies has been intensively 
 	studied \citep[e.g.,][]{saio04,cunha06}. 
 	
 	It has been shown that magnetic coupling 
 	in the layers where the magnetic pressure becomes comparable to the 
 	gas pressure can significantly alter the pulsation frequency. This, in turn, causes non-negligible 
 	deviations from the classical asymptotic theory \citep{tassoul}, which predicts equally spaced 
 	frequencies of consecutive radial overtones and the same degree and is widely used to interpret 
 	high-order p-modes for, e.g., solar-type oscillations. It was shown that the observed 
 	deviations from these spacings depend essentially on the strength of the magnetic field as 
 	well as the field geometry, making accurate frequency determination a valuable tool 
 	in mode identification.
 	
 	The rapid pulsation of 10\,Aql (HD\,176232, HR\,7167, $V=5.91$) was first discovered by \citet{heller} 
 	using high speed photometry with the 1-m telescope of the Mount Laguna observatory. From 
 	about 30\,hrs of observations, they were able to identify three pulsation frequencies at 
 	$f_\mathrm{{1,HK}}=1.4360$\,mHz, $f_\mathrm{{2,HK}}=1.3854$\,mHz and $f_\mathrm{{3,HK}}=1.2393$\,mHz all
 	with amplitudes $<0.5$\,mmag. Due to severe aliasing in their data, and the low 
 	signal amplitudes, they noted that the identifications are not unambiguous; however, they 
 	estimated from the observed spacings that the large frequency separation (hereafter simply 
 	$\Delta\nu$) could be 50.6\,$\mu$Hz. Shortly thereafter, \citet{belmonte91} published 
 	IR photometry which confirmed the detection by \citet{heller} with the signals in their 
 	amplitude spectra being close to the observed frequencies. Surprisingly, their observed amplitudes 
 	(obtained in the $J$ filter) were in some cases about a factor 2--3 greater than in the Johnson $B$ 
 	filter, and their highest peak was coincident with a 1\,d$^{-1}$ alias of $f_\mathrm{{3,HK}}$. One year later, 
 	\citet{belmonte92} returned to 10\,Aql, but failed to redetect the signal by the same method. 
 	
 	IR photometry in the early 1990's constituted the last published photometric campaign of 
 	10\,Aql. Focusing on abundance analysis, \citet{ryab00} obtained 
 	high resolution spectra of 10\,Aql, estimating $T_\mathrm{{eff}}$, radius and $\log g$ through synthetic 
 	line profile fitting. As for many other roAp stars, they detected a strong overabundance of the doubly ionized
 	rare earth elements (REE's) Pr III and Nd III. Shortly thereafter, \citet{kochukhov02} succeeded in 
 	detecting the pulsation from radial velocity variations of these lines while also confirming 
 	the unusually low amplitude of the signal with peak amplitudes of 130-150\,m\,s$^{-1}$. They also 
 	provided the first measurement of the mean magnetic field modulus $\langle B\rangle=1.5\pm0.1$\,kG from 
  Zeeman splitting. \citet{hatzes} confirmed the pulsation in the radial velocities for several lines. 
	They also detected highly variable amplitudes from line to line, but also from night to night for the same line, 
	which was interpreted as either rotational modulation, growth or decay of modes on short timescales, or close 
 	multiperiodicity. As the MOST observations will show, the latter explanation was most likely 
 	the reason.
	
	In an effort to suggest 10\,Aql as a potential target for the space telescope COROT \citep{baglin}, 
	\citet{bigotweiss} published the first and, so far, last attempt to model the oscillation spectrum of this star. 
	Assuming an effective temperature of $8000\pm100$\,K (based on photometric calibrations) and a luminosity of
	$21.4\pm0.6\,L_{\sun}$ (obtained with the Hipparcos parallax), they found an agreement with the 
	three frequencies published by \citet{heller} and the large frequency separation expected for a magnetic 
	field strength of 0.8\,kG.
 	
 	The rotation period of 10\,Aql is unknown; however, it is believed to be very long based on magnetic 
 	field measurements by \citet{babcock}, which showed a reversal of polarity in a timespan of 
 	4 years. \citet{preston} measured a constant mean longitudinal field strength of $\sim500$\,G over 
 	10 consecutive nights. \citet{ryab05} obtained 
 	5 additional values over a timespan of one year, confirming the value measured by Preston some 
 	30 years earlier. Considering the measured $v\sin i=2.0\pm0.5$\,km\,s$^{-1}$ \citep{kochukhov02} as an upper limit, they concluded that the rotational period might possibly be as 
 	long as hundreds of years. Since this, however, would contradict Babcock's observation and a 
 	pole-on configuration could not yet be excluded, there still was no securely measured lower 
 	limit for the rotation period of 10\,Aql.
 	
 	All spectroscopic campaigns conducted on 10\,Aql were based on observations spanning  a 
 	couple of nights and were therefore unsuitable for confirming and precisely identifying the 
 	pulsation modes of the star. Although recent observations have shown the power of 
 	high resolution spectroscopy in probing the interior and even mapping the surface structure 
 	of pulsation in roAp stars \citep[e.g.,][]{kochukhov04}, it is clear that a secure identification of all 
 	pulsation modes is indispensable for a complete understanding of the effects observed (in both 
 	photometry and spectroscopy). Due to the high precision and long uninterrupted time coverage, 
 	MOST is uniquely suited for such a campaign to perfectly complement recent accomplishments 
 	in the field of roAp stars.

%
%________________________________________________________________

\section{Photometry and Data Reduction}

	The MOST (Microvariability and Oscillations of STars) space telescope \citep{walker} 
	was launched in June 2003 to perform high precision uninterrupted photometry of bright stars. 
	It is equipped with a Maksutov telescope with a 15\,cm aperture, a custom broadband filter 
	(350 -- 700\,nm) and, in its original design, two CCD's (one for science, the other for 
	startracking). Due to its sun-synchronous, polar orbit with an orbital period of $\sim 101$\,mins, 
	MOST is able to measure stars for up to two months without interruption, making it 
	uniquely suited to obtain high precision frequency determinations free from aliasing effects. Early in 2006 the tracking CCD system failed due to a 
particle hit. Since then, both science and tracking have been carried out with the 
Science CCD system.
	
	The star 10\,Aql was observed by MOST from June 29 -- July 30 2006 as a primary science target. Because of 
	its moderate brightness ($V=5.91$) it was observed in the open field \citep[corresponding 
	to traditional CCD photometry, see][for details]{walker}. A single exposure consisted of  20 added 0.5\,sec 
	images for a total exposure time of 10\,sec at a sampling interval of 20\,sec. 119724 exposures were acquired with 
	two interruptions due to difficulties with the satellite electronics, to yield a duty cycle of 98\,\%.
	
	\begin{table} 
	\caption{Basic data on the MOST observations of 10\,Aql.}   
	\centering                      
	\begin{tabular}{c c}
	\hline
	\hline
	Parameter		&   Value						\\             
	\hline
	Obs.Time [HJD-2451545]		&   2371.08 -- 2402.24 	\\
	Total timespan [d]			&	31.16				\\
	Duty cycle [\%]				&	98					\\
	Exposure time [sec]			&	10					\\
	Sampling time [sec]			&	20					\\
	\hline                                  
	\end{tabular} 
	\label{tab:obs} 
	\end{table}
	
	Due to parasitic stray light caused mostly by Earthshine scattered into MOST's focal plane, special 
	data reduction techniques were developed to remove instrumental artifacts present in the raw 
	data. In the case of 10\,Aql, three different independent reduction techniques were applied. 
	A newly developed reduction software developed for MOST open field photometry \citep{huber} is 
	based on a technique originally developed for Fabry imaging \citep{reegen06}, in which 
	correlations of intensities of pixels illuminated by starlight and sky intensities 
	are calculated over time. Performing a correction of a linear regression to the intensity-intensity 
	diagrams, stray light effects are effectively eliminated from the data. To be able to apply this 
	technique to open field photometry, the reduction applies an image shifting routine which aligns 
	all images to common pixel coordinates using bilinear interpolation of sub-pixel intensities, 
	allowing a definition of a fixed aperture over time independent of minor satellite jitter effects. 
	The software allows a decorrelation of the mean as well as pixel 
	intensities with the option of using higher order fits (depending on the complexity of the 
	correlation). It additionally implements an identification and correction of pixels affected by 
	cosmic ray hits as well as elimination of  bad quality frames caused by 
	pointing instabilities \citep[see][for details]{huber}.
	
	An independent reduction using the open-field subraster was performed by JR and is described 
	in detail by \citet{rowe}. This technique additionally implements flatfield and 
	dark current corrections. Since there are no moveable parts on the satellite to obtain such 
	calibration frames in the traditional sense, flatfield correction is performed by making use of 
	the fact that during high stray light phases individual CCD subrasters are almost uniformly 
	illuminated by scattered Earthshine. The software uses a combination of 
	PSF fitting and aperture photometry to obtain the count rates of the star. Stray light 
	influences are corrected similar to the reduction described above from a fit to mean 
	star and sky intensities.
	
	With the loss of the MOST startracking CCD another reduction capability was implemented. MOST acquires count rates for several guide 
	stars, which are reduced on-board by simply subtracting a 
	mean sky value after applying a given threshold. Since guide stars and science targets are 
	on the same CCD,  the guide star values are also returned for all science targets (in the case of  10\,Aql, every 0.5\,sec). The extracted on-board processed intensities were additionally reduced by MH 
	by applying the decorrelation technique with other (constant) guide stars. While the accuracy 
	of the count rate determination is certainly not of the same quality as in the actual subraster 
	images, the big advantage of this reduction is that intensities are on-board reduced after each 
	sub-exposure, avoiding possible smearing effects of stray light effects in stacked images.
	
\begin{figure}
   \centering
      \includegraphics[width=0.48\textwidth]{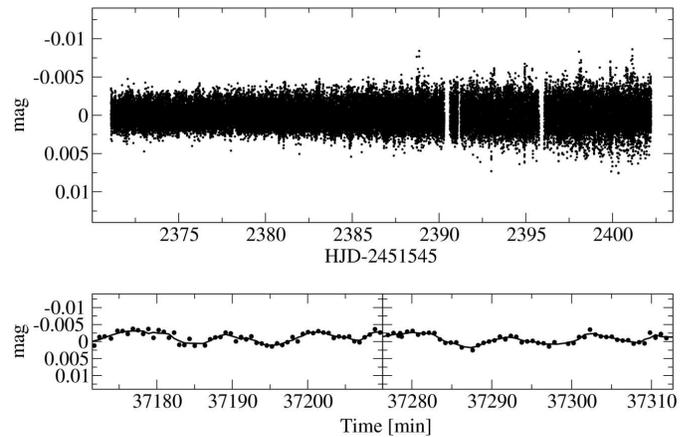}
      \caption{Reduced light curve of 10\,Aql. The bottom 
      panels show a zoom of the light curve, at approximately HJD-2451545 2396.9--2397.0 with a running 
      average overplotted.}
         \label{fig:lc}
   \end{figure}

	All three reductions were performed independently and yielded different effective duty cycles 
	(ranging from 63 -- 95\,\%) depending on the outlier rejection philosophy used. Each reduction 
	showed several strengths and weaknesses, and all three were used to 
	secure the results obtained from the data. Figure \ref{fig:lc} shows the reduced light curve of 
	10\,Aql, in which all signals with frequencies $<5$\,d$^{-1}$ corresponding to instrumental trends have been 
	removed. The data show an increased scatter towards the end of the run. 
	Nevertheless, the bottom panel clearly shows the detection of roAp pulsation.
	
	%
%________________________________________________________________

\section{Frequency Analysis}

	\subsection{Frequency Identification}
	
	Frequency analysis was performed using the programs {\sc SigSpec} \citep{reegen07} as well as 
	Period04 \citep{lenz}. {\sc SigSpec} computes a quantity called spectral 
	significance based on a false-alarm probability by comparing the signal to white noise, 
	and incorporating frequency, amplitude and phase information in the calculation. The program has proved to be extremely powerful for a statistically clean evaluation of the
	signal in Fourier space, and especially in the case of the high frequencies for the roAp stars where 
	colored noise (intrinsic or instrumental) can be assumed to be negligible, the spectral 
	significance gives important additional information compared to the more traditional S/N 
	estimate.
	
	\begin{figure*} 
	\centering 
	\resizebox{\hsize}{!}{\includegraphics{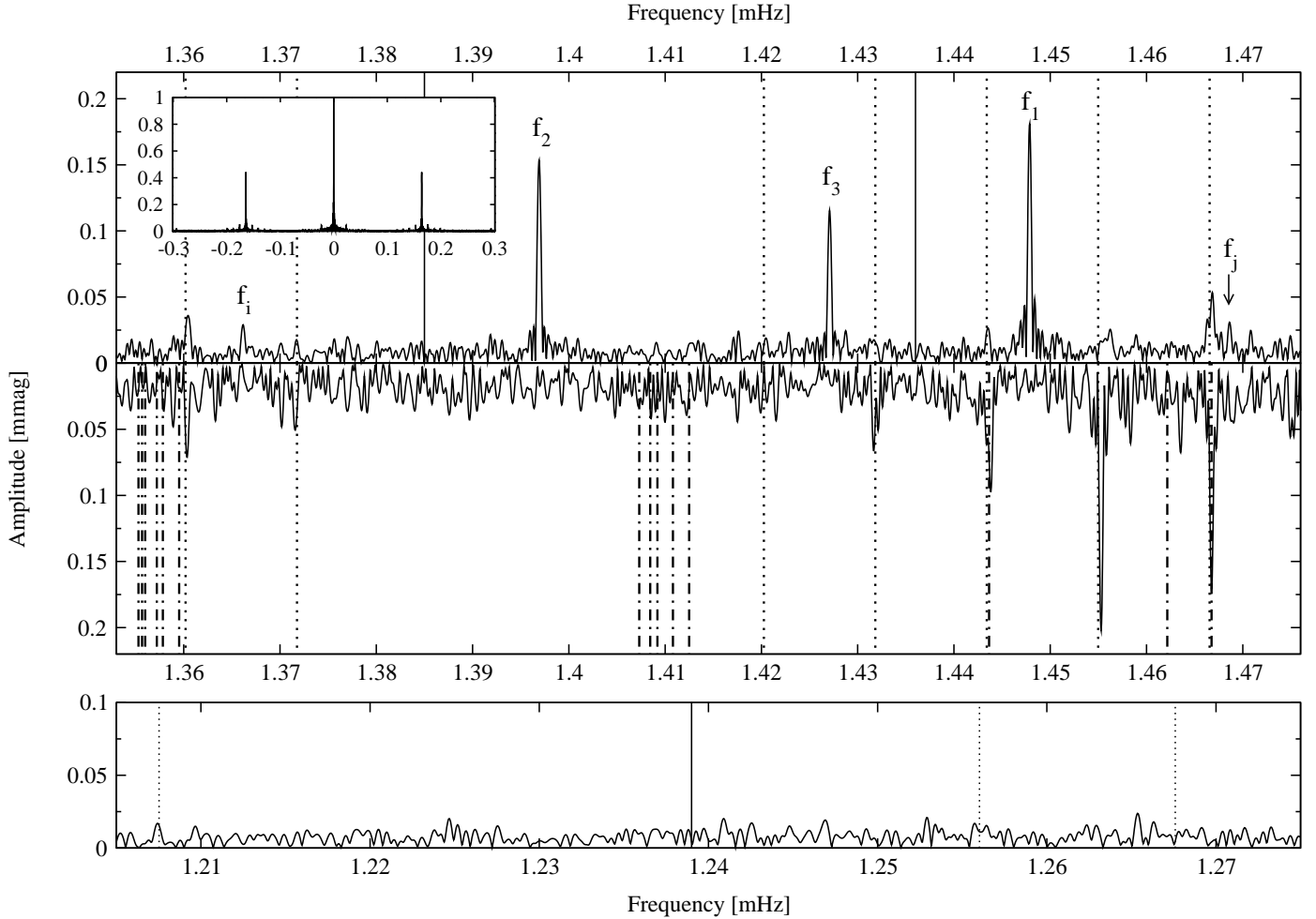}} 
	\caption{(Top panel) Amplitude spectrum of 10\,Aql. Solid lines indicate the position of 
	published frequencies by \citet{heller}. Dotted lines show the position of 1\,d$^{-1}$ sidelobes of harmonics of 
   	MOST's orbital frequency (i.e., instrumental artifacts). The insert displays the spectral window. (Middle panel) 
	Inverted amplitude spectrum of the brightest guide star simultaneously observed 
   	(HD\,175922, $V=7.21$), shown at the same scale. Dashed-dotted lines indicate the position of frequencies 
   	identified in the background readings of the 10\,Aql subraster. (Bottom panel) Frequency region where 
	$f_\mathrm{{3,HK}}$ was reported after prewhitening of $f_{1}$, $f_{2}$ and $f_{3}$ (the position again indicated by a solid line).}
	\label{fig:DFT} 
	\end{figure*} 
	
	Figure \ref{fig:DFT} (top panel) shows a Fourier amplitude spectrum of the reduced 10\,Aql data 
	in the frequency region of interest. With the dashed lines indicating the published frequency 
	values of the two largest amplitude peaks by \citet{heller}, the MOST data clearly show 
	that indeed their frequencies were misidentified as 1\,d$^{-1}$ ($=0.0116$\,mHz) aliases. In fact, for both cases the 
	correct frequency lies 1\,d$^{-1}$ higher. Moreover, the MOST data enables us to resolve an 
	additional, as of yet undetected, frequency around 1.427\,mHz at high S/N. Thanks to the clear 
	spectral window showing only the aliasing effects spaced at MOST's orbital frequency 
	(14.19\,d$^{-1}$ = 0.16\,mHz), 
	the identification of these three frequencies is unambiguous. They appear consistently in all 
	three reductions with comparable significances. To ensure the intrinsic nature 
	of this signal, the middle panel of Figure \ref{fig:DFT} shows the amplitude spectrum of the brightest star 
	simultaneously observed (HD\,175922, $V=7.21$) and additional peaks that were identified using the same approach 
	with {\sc SigSpec} to the light curve constructed with background readings of the 10\,Aql subraster. 
	Clearly, no instrumental power can be identified in the vicinity of $f_{1}$, $f_{2}$ or $f_{3}$. 
	We therefore definitely consider these frequencies to be the result of intrinsic stellar pulsation.
	
	Apart from the three obvious peaks, 10\,Aql does not reveal any eye-catching power above the 
	noise level of about 10\,ppm. We do detect two peaks in the spectrum which were identified 
	by {\sc SigSpec} with a spectral significance of 6, corresponding roughly to a S/N $\ge$ 4. These 
	frequencies are marked in Figure \ref{fig:DFT} by the labels $f_\mathrm{{i}}$ and $f_\mathrm{{j}}$. 
	Unfortunately, these peaks are only identified in the reduction showing the lowest noise level in 
	this region. The comparison data (middle panel, Figure \ref{fig:DFT}) does not reveal any obvious 
	power in this frequency region; however, one must note that the amplitudes of $f_\mathrm{{i}}$ and 
	$f_\mathrm{{j}}$ are at about the noise level of that in the fainter comparison star. Clearly, we are not 
	able to infer from the data available if these peaks are indeed intrinsic or not. Nevertheless, we 
	consider them as candidate frequencies and included them in the subsequent analysis.
	
	Finally, the bottom panel of Figure \ref{fig:DFT} shows the frequency region where an additional 
	mode has been reported by \citet{heller} and \citet{belmonte91}. We do not detect any 
	significant power at $f_\mathrm{{3,HK}}$ or any surrounding 1\,d$^{-1}$ aliases thereof 
	\citep[as suggested by][]{belmonte91}. Considering that $f_\mathrm{{3,HK}}$ was the lowest amplitude 
	frequency in the dataset of \citet{heller}, one might conclude from this result that their 
	detection was indeed an instrumental artifact. Nevertheless, the fact that \citet{belmonte91} also 
	found a signal at almost exactly the same value (in their case, the highest amplitude peak) makes 
	this explanation more unlikely. On the other hand, their follow-up observations one year later did not show 
	evidence for any detectable signal at all using the same methods \citep{belmonte92}. 
	
	If the detections were real, we could conclude that the lack of signal in the MOST data is in fact evidence for a limited mode lifetime 
	on the order of some decades, possibly caused by damping of the mode by intrinsic changes 
	in the star, or a different orientation of the star at the time of MOST observations due to rotation over the 
	past 15 years. Due to the uncertainty of both earlier detections, however, these conclusions must remain in the realm of 
	speculation at this point.
	
	Apart from the frequency region just discussed, the 10\,Aql data do not reveal any 
	variability at lower (or higher) frequencies with a S/N $\ge$ 4 which cannot be 
	attributed to instrumental artifacts. This is consistent with previously published results demonstrating 10\,Aql to 
	be a slow rotator (i.e., no frequency multiplets due to rotational modulation are observed) and theoretical 
	predictions that no low-order p-modes are excited in roAp stars. We are, furthermore, not able to detect 
	harmonics of any definite frequencies $f_{1}$, $f_{2}$ or $f_{3}$, contrary to what has been found for the MOST 
	observations of $\gamma$\,Equ \citep{gruberbauer}. Table \ref{tab:frequencies} lists 
	all identified intrinsic and candidate frequencies in the MOST 10\,Aql dataset. Note that the periods of the 
	stated frequencies correspond to 11.3 -- 12.2 minutes and that errors of the frequency, amplitude and phase 
	values have been computed incorporating the spectral significance as described in \citet{kallinger}.
	
	\begin{table} 
	\caption{List of frequencies identified in the MOST 10\,Aql observations. Candidate frequencies are denoted $f_\mathrm{{i}}$ and 
	$f_\mathrm{{j}}$. The phase values shown are calculated as $\sin(2\pi(f t + \theta))$ 
	with the zero point corresponding to JD 2451545.}   
	\centering                      
	\begin{tabular}{c c c c c}  
	\hline
	\hline
	Id		&   $f$[mHz] 					& 	$A$[mmag] 			&	$\theta$[rad]	& sig \\
	\hline                       
	$f_{1}$	&	1.44786 $\pm$ 0.00003	&	0.17 $\pm$ 0.01 	&	0.2 $\pm$ 0.3		&	135			\\
	$f_{2}$	&	1.39691 $\pm$ 0.00003	&	0.15 $\pm$ 0.01 	&	0.8 $\pm$ 0.3		&	101			\\
	$f_{3}$	&	1.42709 $\pm$ 0.00004	&	0.12 $\pm$ 0.01 	&	0.7 $\pm$ 0.4		&	68			\\
	$f_\mathrm{{i}}$	&	1.3662  $\pm$ 0.0001	&	0.03 $\pm$ 0.02	&	0.3 $\pm$ 1.2	&	6			\\
	$f_\mathrm{{j}}$	&	1.4686  $\pm$ 0.0001	&	0.03 $\pm$ 0.02	&	0.9 $\pm$ 1.2	&	6			\\
	\hline                                  
	\end{tabular} 
	\label{tab:frequencies} 
	\end{table}
	
	\subsection{Amplitude Modulation}
	\label{sec:ampmod}
	
	Apart from mean light variations which are believed to be due to abundance spots on the 
	surface of Ap stars, amplitude modulation of an intrinsic 
	frequency might present clues about the rotation period. Following the oblique pulsator model, 
	the magnetic poles (and hence the pulsation poles) would wander in and out of sight as the star rotates, modulating the pulsation amplitude with exactly 
	that period. As the 31.2 days of MOST observations constitute the longest continuous dataset 
	ever obtained for 10\,Aql, we are able to set a new limit to the so far unknown 
	rotation period of this star.
	
	\begin{figure}
   \centering
  \includegraphics[width=0.48\textwidth]{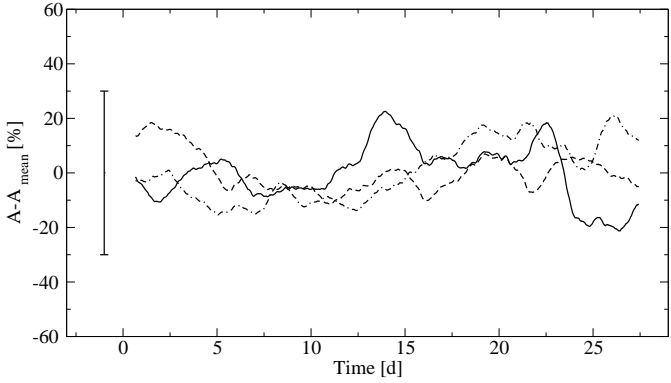}
      \caption{Amplitude modulation of $f_{1}$ as determined by time resolved frequency analysis. 
      The different lines correspond to the three reductions performed for the dataset (solid: 
      Direct Imaging (Vienna); dashed: Direct Imaging (UBC); dashed-dotted: Guide Star), and each 
      line shows a smoothed curve of the individual values. The ordinate is the relative 
      amplitude change. The error bar indicates the mean error for all three reductions 
      for a single datapoint.}
         \label{fig:ampmod}
   \end{figure}
	
	The modulation was investigated by time resolved frequency analysis using {\sc SigSpec}. Subsets 
	were chosen long enough to resolve the frequency component with a secure detection limit (in this 
	case 3.2\,d), with stepwidths equal to one pulsation period. This exercise was performed for all three 
	intrinsic frequencies and all three independent reductions. The results for $f_{1}$ are 
	shown in Figure \ref{fig:ampmod}. While the amplitude does change up to 25\,\% throughout the 
	run, none of the variations are consistent within the different reductions. Similar results 
	have been obtained for $f_{2}$ and $f_{3}$. We conclude that the variations 
	seen are purely due to different data and reduction qualities at different observation times, and 
	that no significant amplitude modulation is present in the 10\,Aql data. Considering the timespan 
	of 31.2 days of observations, we can therefore rule out that the star rotates with a period 
	shorter than about a month (assuming that the star is not seen pole-on, in which case no modulation 
	would be detected even for a rapidly rotating star). This conclusion is also confirmed by 
	time-resolved spectroscopy which was obtained simultaneously to the MOST photometry 
	spanning over 20 days, yielding constant line intensities within 0.5\,\% \citep{sachkov}. 
	Both results are consistent with earlier observations that 10\,Aql is a slow rotator and set a 
	new lower limit for its rotation period.

	%________________________________________________________________

\section{Asteroseismic Analysis}
\label{ana}

	\subsection{Frequency Spacings}
	
	A fundamental observational parameter in stars pulsating with high-order acoustic modes is the 
	large frequency separation $\Delta\nu$, describing the regular spacing of consecutive radial 
	overtones $n$ with same degree $\ell$. Since it is proportional to the square root of the mean density of 
	the star, it allows an estimation of astrophysical parameters such as luminosity and 
	effective temperature.
	
	Obviously, the intrinsic frequencies observed in 10\,Aql 
	do not reveal a regular spacing. The observed spacings are as follows:
	
	\begin{center}
	$f_{1}-f_{2}$ = $\Delta_{1}$ = 50.95 $\pm$ 0.03 $\mu$Hz	\\
	$f_{3}-f_{2}$ = $\Delta_{2}$ = 30.19 $\pm$ 0.03 $\mu$Hz	\\
	$f_{1}-f_{3}$ = $\Delta_{3}$ = 20.76 $\pm$ 0.03 $\mu$Hz	\\
	\end{center}
	
	Considering roAp theory as mentioned in the introduction, this result is not surprising, since it has been shown that 
	the magnetic field can significantly perturb the observed frequencies and therefore the 
	spacing. On the other hand, it might be possible that what we observe are 
	in fact modes of different degree $\ell$. It is, however, interesting to note that 
	$f_{2}-f_\mathrm{{i}}\sim \Delta_{2}$ as well as $f_\mathrm{{j}}-f_{1}\sim \Delta_{3}$. This might 
	give rise to the assumption that either $f_{1}$ or $f_{2}$ are perturbed by the 
	magnetic field, which, if those values were slightly shifted, give four consecutive 
	frequencies spaced by either $\Delta_{2}$ or $\Delta_{3}$.
	
\begin{figure}
   \centering
  \includegraphics[width=0.48\textwidth]{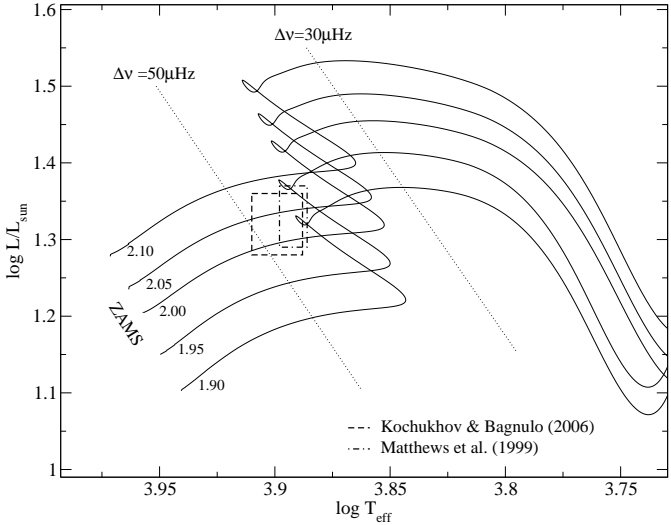} 
      \caption{Theoretical HRD showing evolutionary tracks with $(X,Z)=(0.71,0.19)$ and $\alpha=1.8$ 
      \citep{guenther92}. Numbers next to the tracks show the mass in solar units. The dotted lines 
      indicate contours of constant $\Delta\nu$. Previously determined parameters for 10\,Aql are shown.}
      \label{fig:HRDdeltanu}
   \end{figure}

	To test these speculations, we used the published HRD (Hertzsprung Russell Diagram) position by \citet{matthews} as well 
	as \citet{kochukhov06b}. They calculate the luminosity based on the measured Hipparcos 
	parallax and derive an effective temperature based on photometric calibrations. The frequency spacing, $\Delta\nu$,  
	is directly connected to physical properties which determine the position of the star in the HRD, as given 
	for example by \citet{matthews} via
	
	\begin{equation}
      \Delta\nu = 
      	(6.64 \pm 0.36) \times 10^{-16} M^{0.5} T_{\mathrm{eff}}^{3} L^{-0.75}\,\mathrm{Hz}\:,
  		\label{equ:deltanu}
  \end{equation}
	
	\noindent where $M$ is given in solar masses, $L$ in solar luminosities and $T_{\mathrm{eff}}$ in Kelvin.
	Hence, it is possible to identify models showing a constant $\Delta\nu$ and compare them with 
	the previously determined values. The result is shown in Figure \ref{fig:HRDdeltanu}. In this 
	theoretical HRD, evolutionary tracks using the 
	Yale Stellar Evolution Code \citep{guenther92} are plotted, together with lines of constant 
	$\Delta\nu$ of 50 and 30\,$\mu$Hz, respectively. 
	
	Clearly, any large frequency separation considerably less than 50\,$\mu$Hz would contradict (for the 
	given temperature and luminosity region) the current understanding of roAp stars as objects 
	between ZAMS and TAMS. Note that although these evolutionary tracks do 
	not take into account any effects of the magnetic field, frequency jumps on the order of several 
	tenths of $\mu$Hz are expected to occur only at discrete frequency values rather than constantly 
	for each frequency \citep{cunha06}. More likely, consecutive shifts are on the order of a few $\mu$Hz. It is 
	therefore not possible that the magnetic field perturbations mimic a $\Delta\nu$ which is different
	from that expected from the physical parameters of the star. In other words, equation 
	(\ref{equ:deltanu}) is still a valid approximation even for modes perturbed by a magnetic field, 
	except for large jumps at certain overtones. It is interesting to note that 
	this value for $\Delta\nu$ was actually already suggested after the first photometric campaign 
	by \citet{heller}. Although less accurate and based on wrong frequency identifications, 
	their first asteroseismological interpretation of 10\,Aql might turn out to be correct after all.
	
	\subsection{Model Frequencies}
	\label{sec:modelfrequencies}
	
	To further quantify the observations of 10\,Aql, we performed a comparison with models which were 
	calculated from the zero-age main sequence to the end of the main 
	sequence for masses ranging from 1.90--2.05\,$M_{\sun}$ in steps of 0.05\,$M_{\sun}$ using OPAL 
	opacity tables \citep{iglesias}. In standard models, an initial chemical composition 
	$(X,Z)=(0.70,0.02)$ is adopted and envelope convection is suppressed. Helium abundance is 
	taken as $Y=0.01 + 0.27(x_{2}+x_{3})$, where $Y$ is the helium mass fraction and $x_{2}$ 
	and $x_{3}$ represent fractions of singly and doubly ionized helium, respectively. For standard 
	models, helium is depleted at the surface ($Y_\mathrm{{S}}=0.01$) due to gravitational settling 
	in the outermost layers of the star. Additionally, some non-standard models were computed which included 
	metal-rich composition ($Z=0.025$), envelope convection or assumed no helium depletion near the 
	surface ($Y_\mathrm{{S}}=0.28$). Table \ref{tab:models} lists all the stellar 
	models used in our analysis.
	
	\begin{table}     
	\caption{Summary of models used for the analysis.}
	\centering                           
	\begin{tabular}{c c c c c}                      
		\hline
		\hline
		Id	&  	$M/M_{\sun}$	&	$(X,Z)$		&	He depl		& 	Conv 	\\
		\hline
		S190	&	1.90		&	(0.700,0.020)	&	yes		&	no	\\
		S195	&	1.95		&	(0.700,0.020)	&	yes		&	no	\\
		S200	&	2.00		&	(0.700,0.020)	&	yes		&	no	\\
		S205	&	2.05		&	(0.700,0.020)	&	yes		&	no	\\
		\hline                                  
		H195	&	1.95		&	(0.700,0.020)	&	no		&	no	\\
		H200	&	2.00		&	(0.700,0.020)	&	no		&	no	\\
		H205	&	2.05		&	(0.700,0.020)	&	no		&	no	\\
		\hline                                  
		M200	&	2.00		&	(0.695,0.025)	&	yes		&	no	\\
		\hline                                  
		C195	&	1.95		&	(0.700,0.020)	&	no		&	yes	\\
		\hline
	\end{tabular}  
	\label{tab:models} 
	\end{table}
	
	Theoretical non-adiabatic frequencies of axisymmetric ($m=0$) high-order p-modes under the 
	influence of a dipole magnetic field were calculated using the method described in \citet{saio05}. 
	Outer boundary conditions are imposed at an optical depth of 10$^{-3}$. A reflective mechanical 
	condition ($\delta p/p \to$ constant) was used although for models with $Y_{\mathrm{S}}=0.01$, the 
	acoustic cut-off frequency is close to the frequencies observed for 10\,Aql (for models 
	without helium depletion the critical frequency is higher than the observed frequencies, therefore 
	justifying the use of a reflective boundary condition). Due to the neglect of possible 
	kinetic energy leakage at the outer boundary, and the use of a diffusion approximation for 
	radiative energy transport, the prediction for the stability of the modes in these models may 
	not be very reliable.
	
	Eigenfunctions of theoretical frequencies are expanded in terms proportional to spherical 
	harmonics Y$^{m=0}_{\ell} (\theta,\phi)$. The character of the latitudinal amplitude variation 
	is designated by $\ell_{\mathrm{B}}$, the $\ell$ value of the expansion component having the largest kinetic 
	energy. The value of $\ell_{\mathrm{B}}$ can change as the 
	eigenfunction of a mode is modified by the effects of the magnetic field. In fact, a $\ell_{\mathrm{B}}=0$ 
	mode sometimes changes to a $\ell_{\mathrm{B}}=2$ mode or vice versa. Generally, frequencies of 
	$\ell_{\mathrm{B}}=0,3$ modes are strongly affected by a magnetic field, while 
	$\ell_{\mathrm{B}}=1,2$ modes tend to be less affected. The latter modes can therefore be used 
	to determine the position in the HR diagram and the former modes to estimate the strength of the magnetic field.
	
	Theoretical frequencies were calculated for the HRD position estimated by previous observations 
	as well as the observed frequency spacings and for magnetic field influences ranging from 
	$B_\mathrm{{P}}$ = 0--5\,kG. 

	Because of the necessary simplifications, the models are far from representative of roAp
	stars in detail. For example, the magnetic field is assumed to be a simple centered dipole.
	Consequently, errors of model frequencies needed for the comparison with observational results
	can only be estimated. An additional severe restriction in the models is that only
	axisymmetric modes are considered. In the case of 10\,Aql, however, this assumption can
	be justified by the fact that it has been repeatedly shown that this star has a very long rotation period
	possibly in the order of years, in which case rotationally perturbed modes are negligible. As has been
	shown in section \ref{sec:ampmod}, this is also supported by MOST photometry.

	\subsection{Model Fitting}
	
	\begin{figure*} 
	\centering  
	\resizebox{\hsize}{!}{\includegraphics{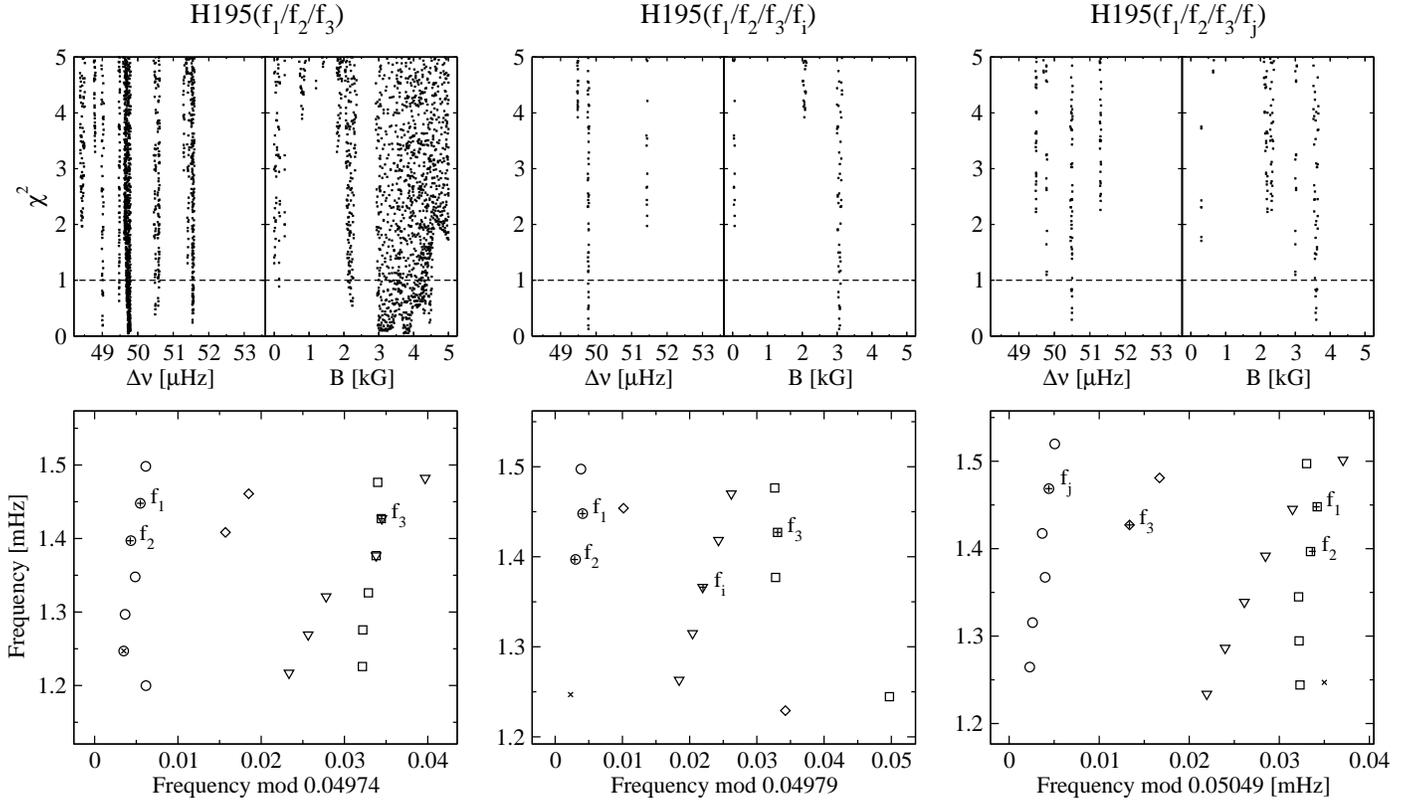}} 
	\caption{Graphical presentation of the model fitting results for different frequency combinations in 
	the H195 model. 
	(Top panels) $\chi^{2}$ values as a 
	function of the parameter space covered by the models in $\Delta\nu$ (i.e., mass, effective temperature and luminosity) 
	and $B_\mathrm{{P}}$, respectively. (Bottom panels) Echelle diagrams of models with the lowest $\chi^{2}$. 
	Spherical degrees $\ell_{\mathrm{B}}=0$ (triangles), $\ell_{\mathrm{B}}=1$ (squares), $\ell_{\mathrm{B}}=2$ (circles) 
	and $\ell_{\mathrm{B}}=3$ (diamonds) are displayed, with observed MOST frequencies labeled and marked by plus signs. 
	The cross in each plot shows the position of a frequency value published by \citet{belmonte91} at 1.247\,mHz. Both 
	errors for observed frequencies and theoretical frequencies are slightly smaller than the symbol sizes.}
	\label{fig:fitresults}
    \end{figure*} 
	
	Due to the fact that the calculation of model frequencies is very time consuming, the 
	resolution of the original grid of models was enhanced by linear interpolation as 
	described by \citet{gruberbauer} in the application to the MOST observations of 
	$\gamma$\,Equ. The stepsize was decreased to 0.02\,kG in $B_{\mathrm{P}}$ as well as 0.00001 in $\log L/L_{\sun}$ 
	and $\log T_\mathrm{{eff}}$.
	
	Observed frequencies were compared to theoretical ones using the $\chi^{2}$ statistics test 
	as described by \citet{guenther04} and defined as
	
	\begin{equation}
		\chi^{2} = \frac{1}{N} \sum_{i=1}^N \frac{(\nu_{\mathrm{obs,i}}-\nu_{\mathrm{mod,i}})^{2}}{\sigma_{\mathrm{obs,i}}^{2}+\sigma_{\mathrm{mod,i}}^{2}}\:,
		\label{equ:chisqu}
	\end{equation}

	\noindent where ${\nu}_{\rm obs,i}$ and ${\nu}_{\rm model,i}$ are the observed and corresponding model 
	frequency, and  ${\sigma}_{\rm obs,i}$ and ${\sigma}_{\rm mod,i}$ are the uncertainties of these 
	values, respectively. $N$ denotes the total number of frequencies fitted. For the uncertainty of 
	model frequencies a value of 0.2\,$\mu$Hz was assumed. Using this formalism, a set of observed frequencies is considered to match the theoretical frequencies within the errors for values of \makebox{$\chi^{2}\leq 1$}. We note that with this formulation we do not draw any conclusions about 
	over- or underestimation of errors (e.g., a reduced $\chi^{2}\ll 1$ value would, in the strictly 
	statistical sense, indicate an overestimation of errors), but rather solely use the relative $\chi^{2}$ 
	values to differentiate between the quality of the fits.
	
	Since the intrinsic nature of the frequencies $f_\mathrm{{i}}$ and $f_\mathrm{{j}}$ cannot be confirmed 
	beyond doubt, model fitting was performed with all combinations of definite 
	and candidate frequencies. The results are summarized in \makebox{Table \ref{tab:fitresults}}. 
	As can be seen from the $\chi^{2}$ values, the observed frequencies fit the 
	theoretical ones well, if not to say too well considering the rather crude magnetic stellar models. 
	Interestingly, when using all five frequencies, no fits within the errors can be found. This naturally 
	leads to the conclusion that, assuming the 
	theoretical frequencies are correct, either $f_\mathrm{{i}}$ or $f_\mathrm{{j}}$ must not be intrinsic to the 
	star. On the other hand, if only the definite frequencies are used, all models fit within the 
	errors, showing that an additional candidate frequency is needed to further constrain the models. 
	Therefore, the focus of attention was brought to the combinations including only one of the 
	two candidate frequencies.
	
	\begin{table}
	\caption{Model fitting results for each combination of definite and candidate frequency set. Bold 
		numbers show the best fitting results for a combination.}
		\centering                      
		\begin{tabular}{c c c c c}                      
		\hline
		\hline
		Model 	& \multicolumn{4}{c}{$\chi^{2}$} \\	
		id		&	\small{($f_{1}$/$f_{2}$/$f_{3}$)}	&	\small{($f_{1}$/$f_{2}$/$f_{3}$/$f_\mathrm{{i}}$)} & \small{($f_{1}$/$f_{2}$/$f_{3}$/$f_\mathrm{{j}}$)} & \small{($f_{1}$/$f_{2}$/$f_{3}$/$f_\mathrm{{i}}$/$f_\mathrm{{j}}$)} \\
		\hline
		S190	&	0.27				&	1.51&	0.94			&	6.35						\\
		S195	&	0.13				&	0.55&	1.33	&	7.96					\\
		S200	&	0.10				&	0.38			&0.69	&	\textbf{2.10}				\\
		S205	&	0.26				&	0.46			& 2.63			&	2.61						\\
		H195	&	0.05	       			&	\textbf{0.13}	       & \textbf{0.29}		&	4.54						\\
		H200	&	\textbf{0.02} &    2.62 &	2.05			&	3.50						\\
		H205	&	0.13				  &	1.40			 & 4.71			&	4.08						\\
		C195	& 	0.11      	&     0.84 &	2.33			&	8.32						\\
		M200	&	0.30		&     5.20			& 4.33			&	15.2						\\	
		\hline
		\end{tabular} 
		\label{tab:fitresults} 
	\end{table}
	
	Although the diversity of $\chi^{2}$ values for all models is quite weak, one can see that 
	(apart from the combination using only definite frequencies), some non-standard models seem 
	to clearly yield worse fits. This is true for models including envelope convection as well 
	as models with higher metallicity $Z$. The former result is consistent with roAp theory, 
	showing that suppressed envelope convection is a necessary condition for driving 
	pulsation in these stars \citep[e.g.,][]{balmforth,saio05} as has also been 
	found by \citet{gruberbauer} for the MOST observations of $\gamma$\,Equ.
	
	Considering only frequency combinations where fits with $\chi^{2} < 1$ can be found, it can 
	be seen that the best fits consistently occur for models without helium depletion. Examining the 
	best fit for the set of definite frequencies (H200), it was found that the corresponding polar magnetic field 
	value lies at 0.7\,kG. While this value seems to be in agreement with the values derived by \citet{bigotweiss} 
	and \citet{kochukhov02}, it must be noted that these authors considered the \emph{mean} magnetic field 
	strength, whereas the pulsation models test the \emph{polar} magnetic field, which is expected to be 
	considerably higher. Therefore, this result would stand in direct contradiction to the observed value inferred 
	by \citet{kochukhov02}. Furthermore, the two primary frequencies are identified as radial ($\ell_{\rm B} = 0$) modes 
	in this model, which contradicts evidence found hitherto for roAp stars as being non-radial pulsators at least for 
	large amplitude frequencies. For these reasons, we conclude that also for the combination of $f_{1}$/$f_{2}$/$f_{3}$ the 
	H195 model, yielding a higher $\chi^{2}$ value (but still $\le 1$) and a polar magnetic field value of $\sim$ 3.7\,kG, 
	presents a more reasonable solution.
	
	In the top panels of Figure \ref{fig:fitresults}, $\chi^{2}$ is shown as a function of $\Delta\nu$ and $B_{\mathrm{P}}$ for the H195 model. To present the results more easily, 
	$T_{\mathrm{eff}}$, $L$ and $M$ have been combined as $\Delta\nu$ according to Equation \ref{equ:deltanu}.
	Including $f_\mathrm{{i}}$, the regions 
	showing low $\chi^{2}$ reduce dramatically (as expected), with only one region in $\Delta\nu$ 
	and $B_{\mathrm{P}}$ having $\chi^{2}<1$. The same is valid for the combination including 
	$f_\mathrm{{j}}$, which, however, does not restrict the parameter space as much as the former combination, 
	with the best fit lying at a slightly higher $B_{\mathrm{P}}$ and higher $\Delta\nu$.
	
	As shown by \citet{saio04} and \citet{saio05}, there are certain values 
	of $B_{\mathrm{P}}$ where the damping of acoustic waves due to magnetic slow waves reaches 
	its maximum. In some cases the expansion of eigenfunctions used in the method does not 
	converge, resulting in gaps in the theoretical frequencies. Note that the region between 
	$B_{\mathrm{P}}$ = 2.3--3\,kG shows such a gap for our set of frequencies. 
	
	In addition to the frequencies observed by MOST, we test the values provided by 
	\citet{heller} and \citet{belmonte91} for the possible (but undetected in MOST data) 
	additional frequency around 1.24\,mHz. Indeed, one of the three values published by \citet{belmonte91} 
	(1.247 mHz) would fit the same sequence of f$_{1}$ and f$_{2}$, as shown in Figure 
	\ref{fig:fitresults} by the cross in the corresponding Echelle diagram. Note that this is not the 
	case for the combinations with $f_\mathrm{{j}}$ and that for the combination using $f_\mathrm{{i}}$, fewer 
	theoretical modes are present since the fit lies very close to the critical $B_{P}$ introducing a gap 
	in the parameter space. Although this result is encouraging, we do not feel comfortable basing a model fit on a 
	frequency that is not observed by MOST. Nevertheless, this is another indication that 
	10\,Aql indeed might have pulsated with an additional, slightly longer period in the early 1990's.
	
%	Analyzing the stability of the identified modes, it is worth noting that, even though the
%	majority of the modes that we do identify are theoretically stable, some modes are also 
%	excited. This is true for example for $f_{1}$ and $f_{3}$ in the configuration using only 
%	definite frequencies, or for $f_{1}$ and $f_\mathrm{{j}}$ including $f_\mathrm{{j}}$. In the solution including 
%	$f_\mathrm{{i}}$, no (identified) modes are found to be unstable. As has however been already mentioned in 
%	Section \ref{sec:modelfrequencies}, the prediction of the stability of modes is not completely 
%	reliable and it has also been found to be in discrepancy with the results for the MOST observations 
%	of $\gamma$\,Equ (Gruberbauer et al.\ 2007).
	It might appear questionable to prefer the quality of one fit over the other when both show a 
	$\chi^{2}<1$, i.e. both solutions fit within the errors. We argue that this is reasonable with the 
	following example: the calculation of the minimum mean deviation of observed to theoretical 
	frequencies, which does not incorporate any error values, gives (e.g., for the combination of 
	certain frequencies) a value of 0.03\,$\mu$Hz for the H195 model. The corresponding standard 
	model S195 (also showing a $\chi^{2}<1$ close to the H195 model, see Table 
	\ref{tab:fitresults}) yields a mean deviation of 0.07\,$\mu$Hz, meaning that the expected values 
	of both theoretical and observed models are closer together for the model showing the lower 
	$\chi^{2}$ value. If we now assume that the error distribution is Gaussian, it is clear that fits with a 
	lower mean deviation from the expected values imply a higher probability that the solution is in 
	agreement with the real (unknown) values.
	
	Despite of this argument, it is clear from the results that there are not enough definite 
	frequencies in the MOST 10\,Aql dataset to obtain a unique solution. Also, it is not possible 
	to clearly rule out whether $f_\mathrm{{i}}$ and $f_\mathrm{{j}}$ are artifacts or 
	not. Indeed, tests have shown that if random values or frequencies detected in simultaneously 
	observed stars with a comparable significance to $f_\mathrm{{i}}$ or $f_\mathrm{{j}}$ are added 
	to the set of definite frequencies, in about 50\,\% of the cases a fit showing a $\chi^{2}<1$ 
	can be found for the H195 model. This result demonstrates the fact that such a small 
	quantity of observed frequencies does not allow one to derive unambiguous results for this star and all conclusions 
	based on these fits should be treated with caution. Nevertheless, it can be concluded from the attempts 
	presented here that the most reasonable fits seem to occur consistently for one 
	stellar model, and that the results including $f_\mathrm{{i}}$ seem to constrain the parameter 
	space better for this model.
	
\begin{figure}
   		\centering
   		\includegraphics[width=0.5\textwidth]{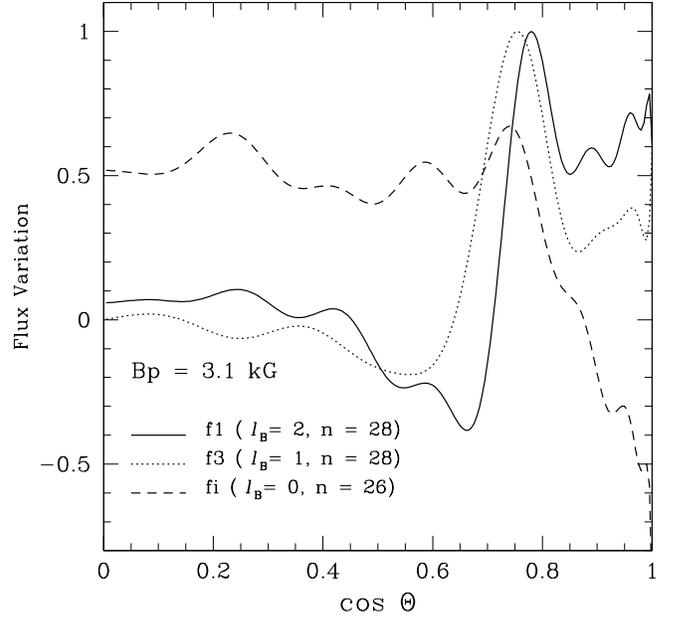}
      	\caption{Flux perturbation as a function of co-latitude $\Theta$ ($\cos \Theta=1$ referring to the poles) 
	for the best fitting non-interpolated H195 model ($\log T_{\rm eff}=3.8883, \log L/L_{\sun}$=1.2473). Note that $f_{2}$ 
	($\ell_{\rm B}=2$, $n=27$) is not specifically shown since the variation coincides with $f_{1}$.}
        \label{fig:fluxpert}
   	\end{figure}

	The bottom panel of Figure \ref{fig:fitresults} shows the Echelle diagrams of the models with 
	the lowest $\chi^{2}$, together with the observed frequencies and a mode identification. 
	Clearly, it is quite dangerous to derive a mode identification considering the low $\chi^{2}$ values given in 
	Table \ref{tab:fitresults} for other models. Nevertheless, we note that for the best solution 
	including $f_{\rm i}$ the mode identification $(\ell_{\rm B}/n)$ would be as follows: $f_{1}=(2,28)$, $f_{2}=(2,27)$, 
	$f_{3}=(1,28)$ and $f_{\rm i}=(0,26)$. If $f_{\rm j}$ is included, the modes would be identified as 
	$f_{1}=(1,28)$, $f_{2}=(1,27)$, $f_{3}=(3,27)$ and $f_{\rm j}=(2,28)$. Figure \ref{fig:fluxpert} shows, 
	for the closest non-interpolated model to our best fit, the flux perturbation as a function of the
	co-latitute $\Theta$ at the phase of maximum light for each mode. It can be seen that for all modes maximum light 
	is concentrated towards $\Theta \sim 0.7$, and that the radial mode exhibits larger amplitudes towards the equator.
	
	\subsection{Comparison with previous results}
	
\begin{figure}
   		\centering
   		\includegraphics[width=0.48\textwidth]{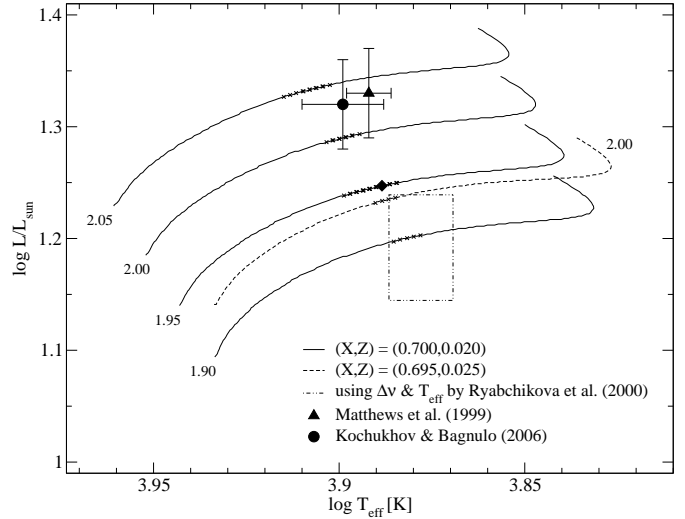} 
      	\caption{Theoretical HRD showing previously determined positions of 10\,Aql using 
      	calibrations based on photometry and the Hipparcos parallax (full circle and triangle) as well as the results 
      	derived in this work. Evolutionary tracks correspond to the models presented in section 
      	\ref{sec:modelfrequencies} (the numbers next to them indicating the corresponding mass). Note that models excluding 
      	He depletion and including envelope convection show no significant deviation from these 
      	tracks. The box shows the position derived assuming $\Delta\nu=50.95$\,$\mu$Hz, using effective 
temperatures derived from spectroscopy and an assumption of the mass to be $1.95\pm0.05$\,$M_{\sun}$. Small crosses show the position of the models listed in Table \ref{tab:models}, with the 
      	filled diamond indicating the best model fit using a combination including $f_\mathrm{{i}}$.}
        \label{fig:result_HRD}
   \end{figure}

	Although the results of fitting are not unambiguous, we can compare our best fitting 
	model with previously determined parameters derived using non-asteroseismological 
	observables. The result of this comparison is shown in a theoretical HRD in Figure 
	\ref{fig:result_HRD}. In addition to the results derived by \citet{matthews} and 
	\citet{kochukhov06b}, we plot the position of our best model fit using the combinations 
	with $f_\mathrm{{i}}$ (filled diamond; note that the combination with $f_\mathrm{{j}}$  only 
	slightly changes this position) as well as an estimate using the observed large frequency spacing 
	in combination with non-asteroseismological parameters by applying equation (\ref{equ:deltanu}).
	The most reliable effective temperature estimation for 10\,Aql is from  \citet{ryab00} 
	who fitted line profiles to the spectral energy distribution, yielding $T_{\rm eff}=7550\pm150$\,K. 
	The mass was assumed to be $1.95\pm0.05$\,$M_{\sun}$.  Using these combined parameters (partially) 
	based on asteroseismological observables, we also derive $R/R_{\sun}=2.3\pm0.2$ and $\log g=3.99\pm0.06$, 
	both values being in agreement with previous determinations \citep{ryab00}. 
	
	There are two notable issues in this comparison. First, one can see that there is 
	a significant discrepancy between the luminosity 
	derived from asteroseismology and Hipparcos. \citet{matthews} were the first to 
	perform a comparison of the parameters derived by both methods for a sample of roAp stars, 
	including 10\,Aql \citep[using the large spacing suggested by][]{heller}. In general, they found a 
	good agreement between the two methods; 
	however, they noted that systematically most Hipparcos luminosities were lower than those ones 
	derived from $\Delta\nu$, \emph{except} for 10\,Aql. Arguing the other way around, they also 
	compared effective temperatures derived using $\Delta\nu$ and the parallax and compared 
	those with temperatures derived from H$\beta$ photometry, concluding again a good 
	agreement but ``In only one case (10\,Aql) does Hipparcos predict a higher temperature.'' 
	If we assume that indeed the measured parallax is inaccurate and 10\,Aql should follow the 
	general trend, then a lower effective temperature and hence a lower luminosity (in agreement 
	with our results) would be needed. Although this explanation is certainly tempting and the 
	deviation noted by \citet{matthews} for 10\,Aql is intriguing, the ambiguous nature of our 
	derived fits makes us cautious about coming to any definite conclusions.
	
	Second, the comparison shows that the effective temperature derivation based on spectroscopy tends to 
	cooler values, while photometric calibrations predict a higher temperature (it has to be 
	noted though that the investigation by \citet{kochukhov06b} was aimed solely to study Ap stars 
	in a consistent manner rather than deriving an exact value for $T_{\rm eff}$). While the asteroseismic value 
	lies in between, it can be seen that only a smaller portion of our models reach into the low temperatures 
	as predicted by spectroscopy (see Figure \ref{fig:result_HRD}), and that our best fitting solution does 
	tend to cooler models, considering the grid investigated in this work. If the observed large frequency 
	separation is indeed correct, such a low value for $T_{\rm eff}$ would present another indication for 
	an overestimated luminosity based on the Hipparcos parallax.
	
	\begin{table}
	\caption{Astrophysical parameters derived for 10\,Aql based on calculations using the 
		observed spacing and model fits (using $f_{\mathrm{i}}$). Bold face numbers show parameters 
		based on estimations and non-asteroseismic observables, respectively. Errors for the model fit have been estimated
		by considering only grid points fits with $\chi^{2} < 1$. The model mass is a fixed 
		parameter.}
		\centering                      
		\begin{tabular}{c c c} 
		\hline
		\hline
		Parameter				&	using $\Delta\nu_\mathrm{{obs}}$	  &	model fit	\\	
		\hline
		$M/M_{\sun}$			&	\boldmath{$1.95\pm0.05$}			&	1.95		\\
		$T_\mathrm{{eff}}$		&	\boldmath{$7550\pm150$}				&	$7734.5\pm0.5$		\\
		$L/L_{\sun}$			&	$15.6\pm1.7$					  &	$17.6885\pm0.0005$		\\
		$B_\mathrm{{P}}$ [kG]	&	-							&	$3.07\pm0.03$		\\
		$R/R_{\sun}$			&	$2.3\pm0.2$					&	$2.3488\pm0.0003$		\\
		$\log g$				&	$3.99\pm0.06$				&	$3.9860\pm0.0001$		\\
		
		\hline
		\end{tabular} 
		\label{tab:results} 
	\end{table}

	%________________________________________________________________

\section{Conclusions}
	
	MOST photometry confirms the rapid pulsation of 10\,Aql, providing for the first 
	time an unquestionable identification of three pulsation modes in this star. Furthermore, we 
	identify two candidate frequencies with too low a S/N for a clear detection. 
	We do not identify any signal at a frequency previously published in two independent ground-based 
	datasets. We, nonetheless, find that this mode fits well with our present solution and 
	conclude that it might be evidence for finite mode lifetimes in roAp stars, 
	if the detections are real. Our results confirm previous findings that 10\,Aql pulsates 
	with very low amplitudes, and that if additional modes are present their amplitudes must lie 
	significantly below a level of about 40\,ppm.
	
	Based on the detected signal we can conclude that previously detected amplitude modulation 
	in spectroscopy by \citet{hatzes} was most likely due to beating of unresolved 
	intrinsic frequencies. This is confirmed by simultaneous spectroscopic monitoring, which shows that 
	the observed amplitude modulation in 10\,Aql is a pure beating effect \citep{sachkov}.
	
	The MOST light curve presents the longest continuous dataset ever obtained for this star. 
	A time-resolved analysis of the amplitudes for the detected frequencies yields no significant 
	modulation. We conclude that in case the star is not seen pole-on, 10\,Aql must rotate on 
	timescales longer than a month, confirming previous findings \citep[e.g.,][]{ryab05}. This is 
	supported by the measured $v\sin i=2.0\pm0.5$ \citep{kochukhov02}, which, 
	together with the radius derived in this work, yields an approximate lower limit for the inclination 
	of $i>30\pm10^{\circ}$.
	
	Our attempt to fit the observed frequencies to theoretical models does not yield an 
	unambiguous solution. We note, however, that models including envelope convection or higher 
	metallicity give a poorer fit. For all frequency combinations fitting within the errors, a 1.95\,$M_{\sun}$ model 
	without helium depletion yields the most reasonable solution. The derived \emph{polar magnetic field} strength is about twice the 
	value previously published for the \emph{magnetic field modulus}, a result that reappears for  
	the MOST observations of $\gamma$\,Equ \citep{gruberbauer}. The derived 
	position of 10\,Aql deviates by less than 2\,$\sigma$ from the luminosity based on the Hipparcos 
	parallax. We conclude that this might be due to an inaccurate parallax measurement, 
	but caution that our results must be considered with care as discussed in section \ref{ana}. 
	We also note that the spectroscopic data obtained simulaneously with MOST shows 
	evidence for a pulsation node at a smaller optical depth (i.e., higher in the atmosphere) than considered in the 
	current models \citep{sachkov}. Therefore, the modeling effort and derived results presented in this work must be 
	considered as only a first attempt to understand the pulsation of this star. Still, our results presented here also 
	reflect state of the art modelling efforts for pulsating magnetic stars and indicate the strong need for improved 
	theoretical concepts.
	
	Despite the most complete and precise photometric campaign ever obtained for this star, the 
	oscillation spectrum of 10\,Aql remains (compared to other, better studied members such 
	as $\gamma$\,Equ or HR\,1217) poorly understood. Nevertheless, the MOST observations have certainly 
	shed some new light on this star and provided new insights that might help to unravel 
	the mysteries of this roAp star in future observations and modeling efforts.

	%________________________________________________________________

	\begin{acknowledgements}
    We are thankful to Mikhail Sachkov, Tanya Ryabchikova, Oleg Kochukhov and Nicole Nesvacil 
    for discussion and collaboration with simultaneous spectroscopy.
    It is also a pleasure to thank Chris Cameron for valuable discussion on the theory of frequency 
    spacings in roAp stars. DH, MG, MH, WWW, TK, PR and RK are 
    supported by the Austrian Science Fund (FWF P17580). The Austrian participation in the MOST 
    project is funded by the Austrian Research Promotion Agency (FFG). HS is 
    supported by the 21st Century COE programme of MEXT, Japan. DBG, JMM, AFJM, and SR acknowledge 
    funding from the Natural Sciences \& Engineering Research Council (NSERC) Canada. 
    This research has made use of the SIMBAD database, operated at CDS, Strasbourg, France.
	\end{acknowledgements}
	
	\bibliographystyle{aa}
	\bibliography{9920}
	
	\end{document}